# All Inkjet-printed Organic Solar Cells on 3D Objects

Marc Steinberger[1,*], Andreas Distler[1], Johannes Hörber[3], Kai Cheong Tam[1], Christoph J. Brabec[1,2,*], and Hans-Joachim Egelhaaf[1,2]

**Abstract**

Drop-on-demand inkjet printing is a promising and commercially relevant technology for producing organic electronic devices of arbitrary shape on a wide variety of different substrates. In this work we transfer the inkjet printing process of organic photovoltaic devices from 2D to 3D substrates, using a 5-axis robot system equipped with a multi-nozzle inkjet printing unit. We present a ready-to-use 3D printing system for industrial application, using a 5-axis motion system controlled by commercial 3D motion software, combined with a commonly used multi-nozzle inkjet print head controlled by the corresponding printing software. The very first time inkjet-printed solar cells on glass/ITO with power conversion efficiencies (PCE) of up to 7% are realized on a 3D object with surfaces tilted by angles of up to 60° against the horizontal direction. Undesired ink flow during deposition of the inkjet-printed layers was avoided by proper ink formulation. In order to be able to print organic (opto-)electronic devices also on substrates without sputtered indium tin oxide bottom electrode, the bottom electrode was inkjet-printed from silver nanoparticle (AgNP) ink, resulting in the first all inkjet-printed (i.e., including bottom electrode) solar cell on a 3D object ever with a record PCE of 2.5%. This work paves the way for functionalizing even complex objects, such as cars, mobile phones, or "Internet of Things" (IoT) applications with inkjet-printed (opto-)electronic devices.

## 1 Introduction

Direct functionalization of objects with arbitrary shape is necessary in order to access cheap and versatile production of smart functional devices (1–5). To enable small complex devices to be sustainably smart and functional and e.g. be part of an "Internet of Things" (IoT) network, local energy harvesting is needed (6). This can be achieved by attaching solar cells to the device, which may, however, lead to higher costs, more waste, and design limitations for the device. Methods to directly print solar cells onto a device, like drop-on-demand inkjet printing, are able to overcome these issues (7). Finally, combining direct inkjet printing with a multi-axis robot system enables direct application of solar cells even onto complex 3D objects.

Organic solar cells (OSCs) already attracted attention in research and industry because they are lightweight, semi-transparent, mechanically flexible, and compatible with high-throughput production methods like slot-die coating, gravure printing, spray coating, and inkjet printing, which enables an economical, fast, and versatile production (8–11). Due to the recent development of novel photoactive materials, OSCs now reach up to 19% power conversion efficiency (PCE), demonstrating its competitiveness against other PV technologies (11–22).

Inkjet printing is a technology that is widely used in industry and research because it is a contactless and high-resolution digital printing method that allows efficient use of material, high throughputs, and low-cost manufacturing. Therefore, it is also the printing technology of choice for printing organic photovoltaics (OPV) of arbitrary shapes.

Inkjet printing of a complete OSC stack (i.e., all functional layers including both electrodes) has already been successfully demonstrated (1). While inkjet printing of the charge extraction layers (ETL and HTL) as well as the top electrode was found to be relatively straightforward, the active layer (due to its morphological constraints) and the bottom electrode (due to shunt-causing agglomerates) pose a bigger challenge. Further optimizations and novel active layer materials such as non-fullerene acceptors (NFAs) helped to improve the PCE of inkjet-printed solar cells to 9-16%,

reaching 15.78% PCE as recently reported by Sang et al.(2,23–32)

However, the fabrication of inkjet-printed solar cells on 3D objects has not been achieved so far.(33) In this work we demonstrate inkjet printing of OSCs on 3D objects by controlling the morphology and topology of all five fuctional layers, including the bottom electrode. In addition, we discuss the major challenges and provide solutions that will facilitate the industrialization of this technology.

## 2 Methods

### 2.1 Materials

The alcohol-based silver nanowire (AgNW) formulation ClearOhm was obtained from Cambrios. The hole-transport layer (HTL) material, poly(3,4-ethylenedioxythiophene) polystyrene sulfonate (PEDOT:PSS) was purchased from Heraeus. Livilux Super Yellow (PDY-132) from Merck, PV2000 from Raynergy Tek, and $PC_{61}BM$ from Solenne BV were used as active materials. Zinc oxide (ZnO) ink from Avantama (N-10) with 2.5 wt% concentration was used for the electron transport layer (ETL). As bottom electrode, silver nanoparticle (AgNP) paste from Dowa Electronics Materials Co. Ltd was used. Pre-patterned indium tin oxide (ITO) glass (25 mm x 25 mm x 1 mm) was obtained from Liaoning Huite Photoelectric Technology Co., Ltd.

### 2.2 Substrate preparation

Inkjet printing was performed on four different substrates. 2D solar cells were printed on glass/ITO and pure glass substrates. 3D solar cells were deposited on glass and glass/ITO substrates attached to a 3D structure at four different angles with respect to the print head, namely 0°, 30°, 45°, and 60° (see Figure 1a). Before printing, all substrates were cleaned by ultrasonication in acetone and 2-propanol for 5 min each.

### 2.3 Device fabrication process

For inkjet printing of AgNP bottom electrodes on glass substrates, wetting was improved by coating a thin layer of PEDOT:PSS (AI4083 diluted 1:5 with IPA and filtered using a 0.45 µm PTFE filter) beforehand.

The formulation of inkjet printing inks was done according to the following recipes. AgNP paste was dissolved in 2-butoxyethanol at a concentration of 29 wt% and afterwards filtered three times using 1 µm, 0.45 µm, and 0.45 µm filters consecutively. ZnO ink was produced using commercially available N10 and replacing the solvent (2-propanol) with 1-pentanol at same concentration by means of rotary evaporation. The reformulated ink was filtered using a 0.22 µm PVDF filter. $PV2000:PC_{61}BM$ (1:1.5) solution was prepared using a total concentration of 37.5 mg/ml in o-xylene:tetralin (1:1). The solution was stirred over night at 120 °C in nitrogen atmosphere and cooled down to room temperature before printing. The PEDOT:PSS ink was produced by filtering a 1:1 mixture of Clevios FHC Solar by Heraeus and purified water using a 0.45 µm PVDF filter. Afterwards, the fluorosurfactant Capstone FS-31 was added to the ink with a concentration of 5 µl/ml. The mixed ink was placed in the ink reservoir three hours prior to the printing process to avoid bubbles. The AgNW ink was obtained by replacing the original solvent (2-propanol) with 1-pentanol at one third of the original concentration. Before printing, the reformulated ink was filtered using a 30 µm filter.

The printing parameters are summarized in Table 1. Spectra S-class inkjet print heads were used for all experiments. All layers were printed with a resolution of 550 dpi. AgNP layers were sintered at 160 °C for 10 minutes, ZnO layers were annealed for 5 min at 140 °C in air. PEDOT:PSS and AgNWs were annealed for 4 min under nitrogen at 140 °C and 120 °C, respectively.

*Table 1: Inkjet printing parameters of different inks for Spectra S-class print heads using piezoelectric voltages between 100-120 V and a piezo-signal with ramp up, hold, and down times between 1-25 ms.*

| Ink | Voltage [V] | Pulse up [ms] | Pulse hold [ms] | Pulse down [ms] |
|---|---|---|---|---|
| AgNP | 120 | 4 | 9 | 10 |
| ZnO | 100 | 6 | 10 | 6 |
| PV2000: PCBM | 100 | 1 | 8 | 1 |
| PEDOT:PSS FHC | 120 | 15 | 15 | 1 |
| AgNW | 120 | 25 | 25 | 1 |

### 2.4 3D inkjet equipment

For inkjet printing on 3D substrates a special inkjet printer was built and its control unit and software developed or adapted (see Figure 1b+c). A 5-axis robot system was equipped with a Spectra S-class inkjet print head and controlled by Meteor software and hardware. The speed of the different axes was transformed into transistor–transistor logic (TTL) signals that feed the printing software with the required printing frequency according to the resolution. Complex 5-axis movements were then programmed in the Software Motion 3D using CAD models of the respective object to print on.

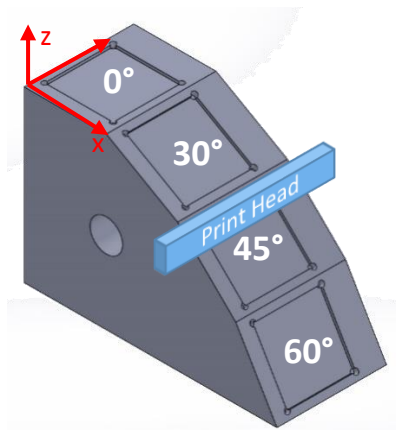

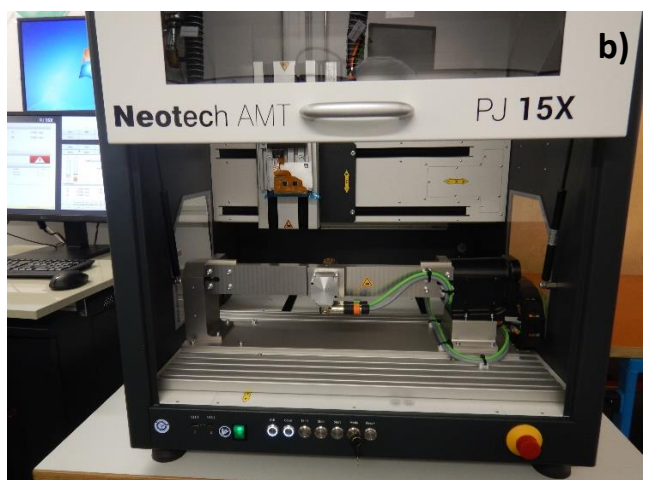

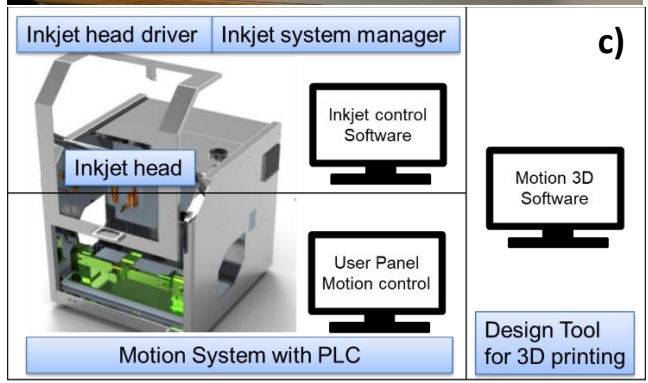

*Figure 1: a) 3D object holder with inkjet printing unit above. B) 5-axis robotic system with inkjet printing unit. C) Schematic of the 3D inkjet printing control units.*

*2.5 Characterization:*

For analysing the topology and surface properties of the layers, an optical microscope from Olympus and a confocal microscope from NanoFocus were used. The solar cell efficiency was measured using an AAA solar simulator from LOT Quantum Design providing 1000 W/m² (AM1.5G).

## 3 Results and discussion

For investigating inkjet-printed solar cells on 2D and 3D objects, the functional layers of the organic photovoltaic (OPV) layer stack were inkjet-printed either onto glass/ITO substrates or onto pure glass substrates. In the latter case, the bottom electrode was provided by inkjet-printing a silver nanoparticle (AgNP) layer (see Figure 2).

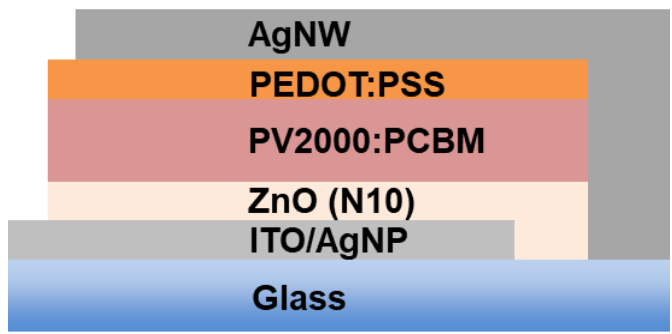

*Figure 2: Schematic layer stack of all inkjet-printed solar cells with either ITO or inkjet-printed AgNP bottom electrode.*

### 3.1 3D inkjet-printed solar cells with ITO bottom electrode

For the development of a printing process for 3D-printed solar cells, a special 3D object with differently tilted surfaces at angles of 0°, 30°, 45° and 60° with respect to the horizontal (see Figure 1a) was designed and is mounted underneath the print head of the robotic 5-axis inkjet printing system. In order to investigate the effect of different impact angles of the inkjet droplets on layer formation and functionality, the printhead is moved using only three axes (x,y,z) during this experiment to ensure constant drop impact angles. Using this setup, complete organic solar cells of the stack ZnO/PV2000:PCBM/PEDOT:PSS/AgNW are then inkjet-printed onto glass/ITO substrates at these four different angles. Figure 3 shows the resulting power conversion efficiencies (PCE) of the devices printed at different inclination angles of 0° (grey), 30° (red), 45° (green) and 60° (blue). The results show similar performances for all variations with maximum PCE values between 5.6-6.5% PCE. The solar cells printed at 30° and 45° show a slightly lower performance, which is probably only related to statistical variations. The best solar cell with 6.5% PCE as well as the highest average performance is achieved at a printing angle of 60°.

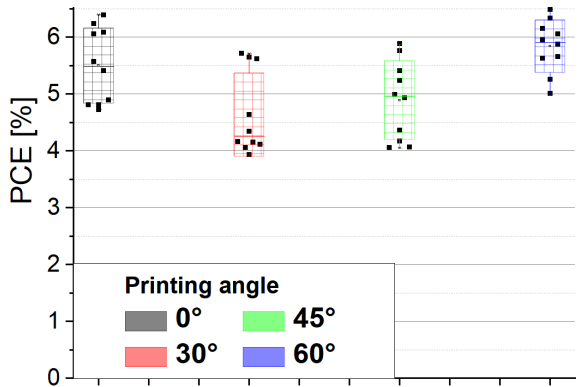

Figure 3: 3D inkjet-printed solar cells on glass/ITO substrates with the layer stack ZnO/PV2000:PCBM/PEDOT:PSS/AgNW prepared at different inclination angles between print head and substrate.

## 3.2 3D inkjet-printed solar cells with printed bottom electrode

Since printing solar cells onto pre-coated ITO substrates limits the choice of substrates and therefore also possible applications, an injekt printing process for silver nanoparticle (AgNP) bottom electrodes is developed, which enables the fabrication of fully printed organic electronic devices (OEDs) onto any object.

### 3.2.1 Inkjet printing of AgNP electrodes

Besides a high conductivity, the most important requirement for bottom electrodes of OEDs is a low roughness, as too rough surfaces may pierce through the overlying layers and, by this, short-circuit the whole device. Figure 4a depicts a confocal microscope image of an unsuitable (i.e., too rough) AgNP electrode that has been inkjet-printed with a non-optimized printing process. A multitude of spikes is observed leading to a high overall roughness with a root mean square height of Sq = 0.034 µm.

In order to obtain smooth AgNP layers, we optimize the printing process by investigating the influence of printing resolution and wet film drying temperature on the homogeneity of the layer. To this end, different AgNP layers are printed with 400-700 dots per inch (dpi) and subsequently dried at room temperature, 40 °C, 65 °C, or 90 °C. Afterwards, all layers are thermally annealed for 5 min at 180 °C.

The results of this optimization are shown in Figure 4b by plotting the roughness (root mean square height) versus printing resolution for different pre-drying temperatures. Room temperature drying of the wet film results in the overall highest roughnesses with Sq values of up to 0.032 µm for 700 dpi. All other pre-drying temperatures between 40 °C and 90 °C lead to significantly lower roughnesses independently of the printing resolution. Wet film drying at 65 °C shows the smoothest surfaces between 0.014-0.015 µm, the lowest roughness being obtained for a printing resolution of 700 dpi.

The respective sheet resistances of all samples are plotted in Figure 4c. The observed trend of decreasing resistance with increasing resolution is valid for all pre-drying temperatures and can be explained by an increasing amount of AgNP deposition per area at higher dpi values. The sheet resistance is found to be generally independent of the wet film drying temperature and shows values below 1 Ω/sq for all samples printed with a resolution of 500 dpi or higher. Decreasing the temperature of the subsequent thermal annealing from 180 °C to 160 °C increases the sheet resistance to the 2-3 Ω/sq range, which is still sufficient for the application as bottom electrode for OSCs, without significantly changing the surface roughness properties (see Figure S1).

To conclude, the best parameters for the inkjet printing of AgNP bottom electrodes are 700 dpi resolution, 65 °C pre-drying, and 5 min post-annealing at 160 °C. These parameters will be used in the following for the fabrication of all inkjet-printed solar cells.

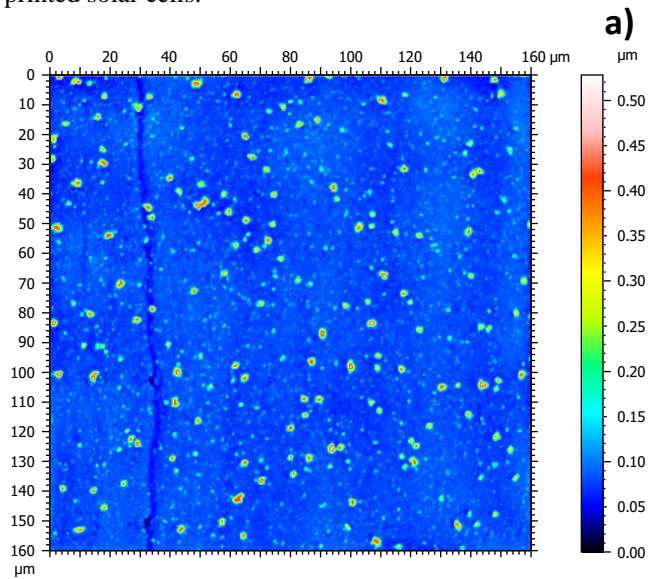

a)

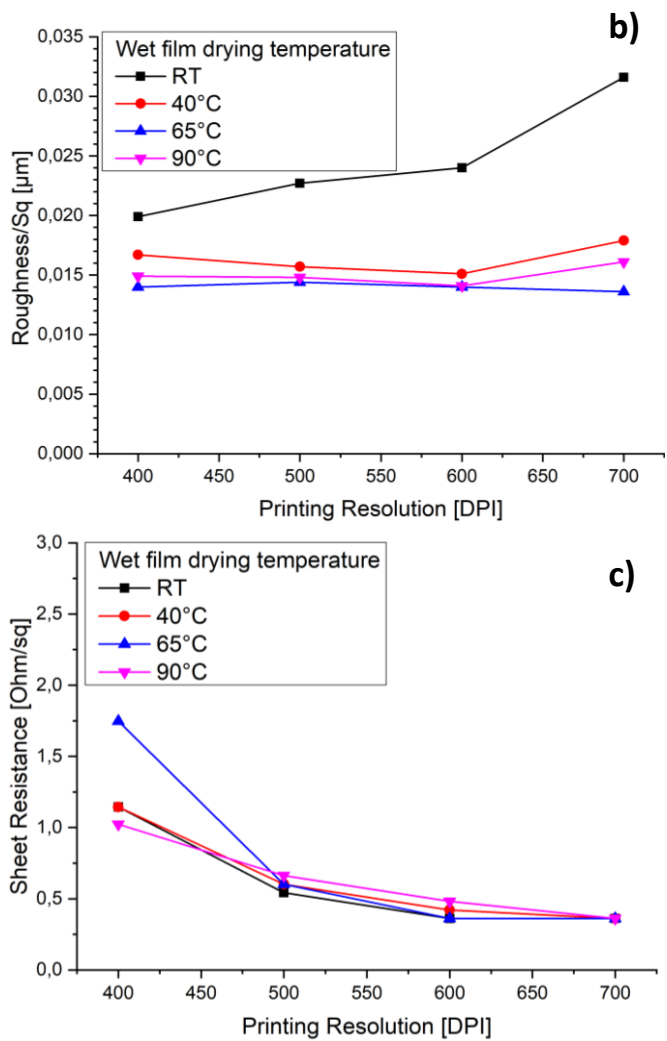

*Figure 4: a) Confocal microscope image of a AgNP film on glass inkjet-printed with 700 dpi resolution, pre-dried at room temperature and annealed at 160°C for 5 min. Root mean square height (Sq) (b) and sheet resistance (c) of AgNP layers on glass inkjet-printed with different resolutions, pre-dried at different temperatures, and subsequently annealed at 180 °C for 5 min.*

### 3.2.2 All inkjet-printed solar cells

Using the optimized process for printing AgNP bottom electrodes, fully inkjet-printed OSCs are processed onto glass substrates, first at 0° angle, in order to optimize the layer stack and to further investigate the influence of the distance between print head and substrate, since this parameter will always vary to a certain extent in a multi-nozzle inkjet printing process on 3D objects. The electron transport layer (ETL) is deposited at three different thicknesses by printing either one, two, or three 60 nm thick layers of zinc oxide consecutively. This is alongside a systematic variation in the distance from printhead to substrate between 0.5 mm (standard) and 3.5 mm. Like this, the effect of bottom electrode roughness on the device performance can be investigated, since thicker ETL layers are able to cover a rough surface better and can thus prevent shunting of the device to a certain extent.

The resulting solar cell performances are displayed in Figure 5a and show two general trends: firstly, an increasing efficiency with increasing ETL thickness, and secondly a decreasing efficiency for higher printing distances. On top the efficiency is reduced compared to cells printed on glass/ITO, due to shunting of the rough AgNP layer through the layer stack. The combination "highest printing distance and thinnest ZnO layer" leads to the worst performance among all variations and increasing the ZnO thickness shows the strongest positive effect for this printing distance. The fact that for all printing distances the PCE increases with the ZnO layer thickness indicates that the roughness of the bottom electrode causes shunting of the devices which limits device performance.

While there is no difference in performance observed between the standard printing distance of 0.5 mm and 1.5 mm, the PCE already slightly decreases at a distance of 2.5 mm and becomes significantly worse at 3.5 mm, suggesting that at higher printing distances the roughness of the bottom electrode increases. This is confirmed by the microscope images of AgNP dots that are inkjet-printed with different distances between print head and substrate (Figure 5b). One can clearly see that the quality of the dots decreases with increasing printing distance. While the appearance of a drop printed with 1.5 mm distance is still similar to one that is printed with the standard printing distance of 0.5 mm, increasing the distance further leads to a decrease in dot diameter and a frayed edge. Furthermore, a "coffee ring effect" and a generally increased layer inhomogeneity is observed once the printing distance becomes too high. Therefore, the distance between nozzle and substrate must be as small as possible and should not exceed 1.5 mm.

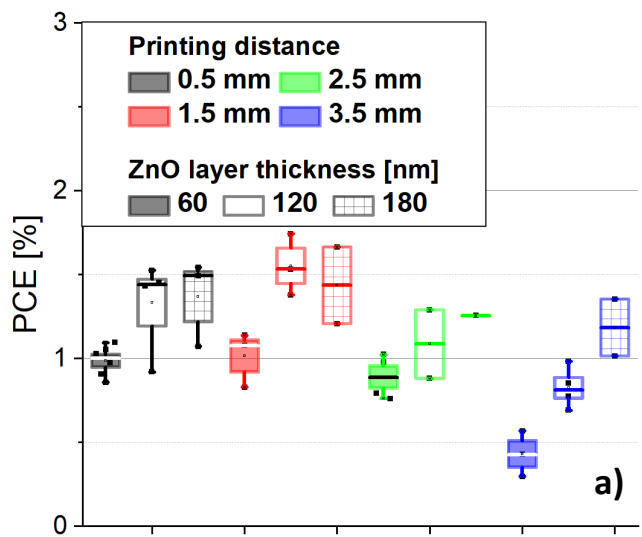

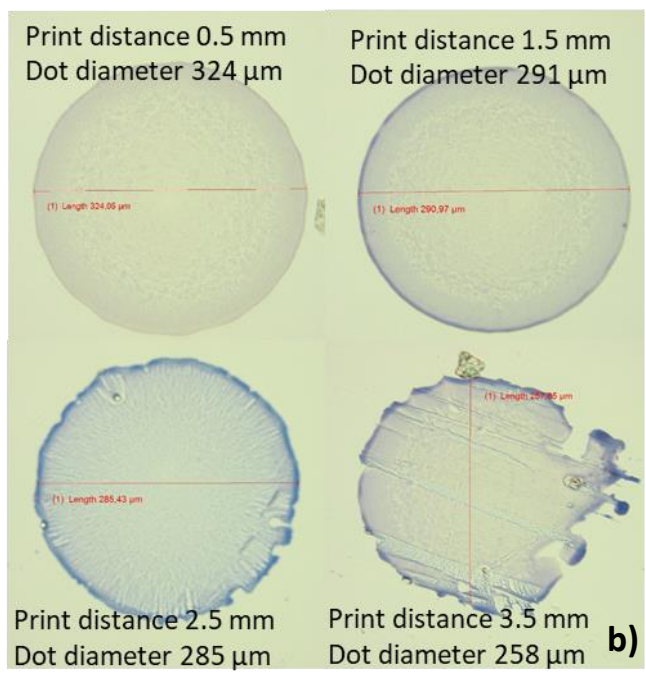

*Figure 5: a) Power conversion efficiencies of all inkjet-printed solar cells on glass substrates with the layer stack AgNP/ZnO/PV2000:PCBM/PEDOT:PSS/AgNW having different layer thicknesses of ZnO and having been printed with different distances between print head and substrate. b) Microscope images of AgNP dots inkjet-printed with different distances between nozzle and substrate.*

Combining all of the developments described above, finally all inkjet-printed solar cells with the same layer stack are printed onto the previously described 3D object with different surface angles of 30°, 45°, and 60°. The resulting PCE values of the photovoltaic devices are plotted in Figure 6a. Efficiencies of up to 3.0% are achieved for 30°-printed solar cells, while 45°- and 60°-printed solar cells showed maximum PCE values of 1.8% and 2.5%, respectively. The general trend that higher printing angles lead to lower average PCEs can be correlated with the observation that the AgNP wet film starts to flow at high angles (see Figure 6b). For the AgNP ink this effect is already observed at 30° to a certain extent and becomes more pronounced at higher angles, whereas none of the other functional layers showed this behavior, even at 60° printing angle.

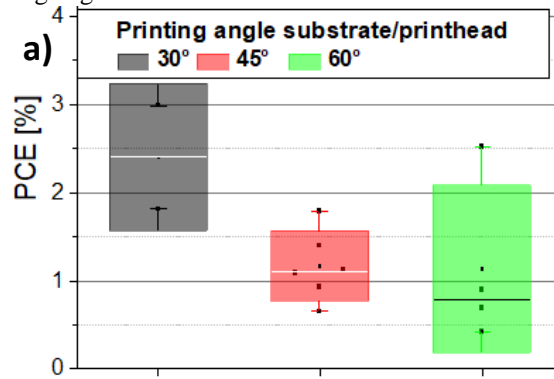

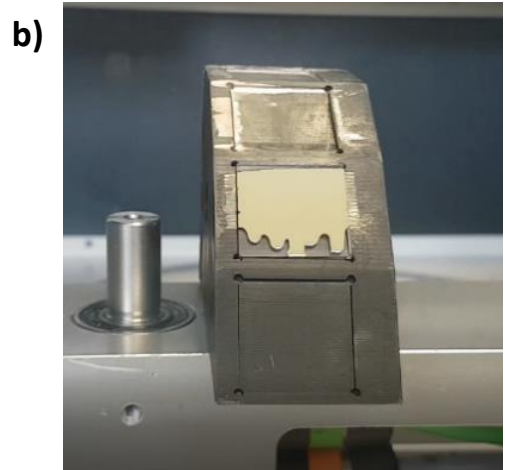

*Figure 6: a) Power conversion efficiencies of all inkjet-printed solar cells with the layer stack of AgNP/ZnO/PV2000:PCBM/PEDOT:PSS FHC/AgNW printed on a 3D object with glass surfaces of different angles with respect to the print head. b) Wet film of AgNP ink on glass inkjet-printed with an angle of 45°.*

## 4 Conclusion and Outlook

In this work we developed a 3D inkjet printing process for organic electronic devices by systematically investigating the influence of printing distance and angle on the functional layers. A commercially available 5-axis robot system equipped with a multi-nozzle inkjet print head was used to produce the first 3D-printed organic solar cells. Printing of organic solar cells onto a 3D object with glass/ITO surfaces has been shown to result in high power conversion efficiencies up to 6.5%, even for high inclination angles of 60° between print head and substrate. Complete inkjet printing of the whole solar cell stack including a AgNP bottom electrode onto a 3D object was also achieved and yielded PCE values up to ~3%. In this case, material flow of the AgNP ink on tilted surfaces and the resulting bottom electrode roughness was found to limit device performance. Once this open challenge is overcome, e.g. by further improving the ink formulation, inkjet printing of highly efficient solar cells directly onto 3D objects will be possible.

Apart from organic solar cells, also other optoelectronic devices such as organic light-emitting diodes (OLEDs) can be printed using the setup and the processes presented in this work, as shown in Figure 7, which makes this technology very versatile and industrially relevant.

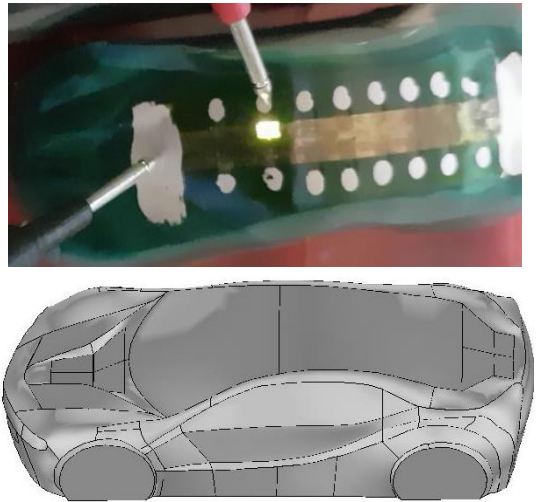

*Figure 7: All inkjet-printed OLED with the stack AgNP/ZnO/Super Yellow/PEDOT:PSS FHC/AgNW printed onto a 3D model of a car.*

## 5   Acknowledgements

DOWA is acknowledged for providing the silver nanoparticle ink. The authors acknowledge the "Solar Factory of the Future" as part of the Energy Campus Nürnberg (EnCN), which is supported by the Bavarian State Government (FKZ 20.2-3410.5-4-5). The Federal Ministry for Economy and Climate is acknowledged for financial support of the IGF project "OLE3D" (19854 N). H.-J. E. and C. J. B. acknowledge funding from the European Union's Horizon 2020 INFRAIA program under Grant Agreement No. 101008701 ('EMERGE'). Part of this work has been supported by the Helmholtz Association in the framework of the innovation platform "Solar TAP".

**Supplementary information**

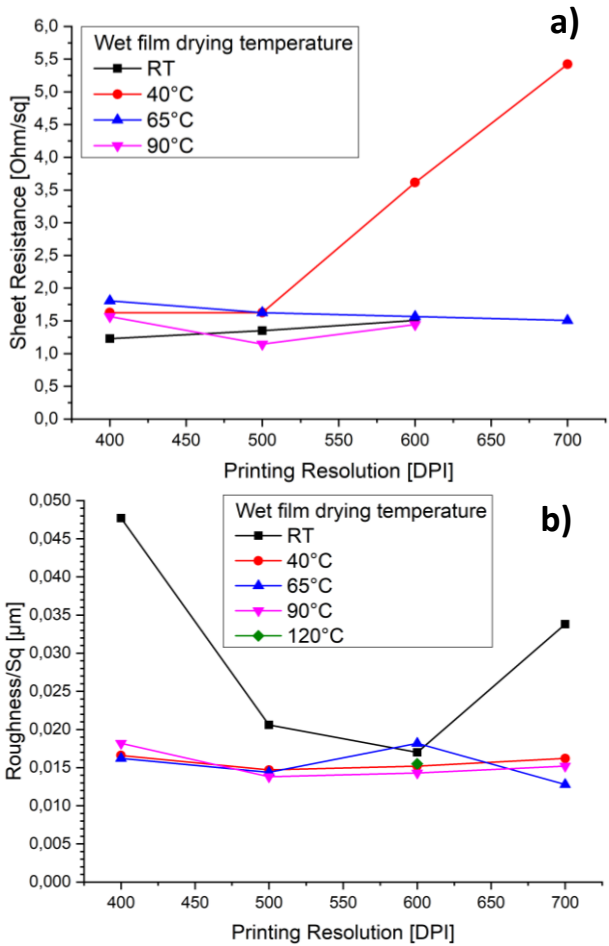

*Figure S1: Root mean square height (Sq) (a) and sheet resistance (b) of AgNP layers on glass inkjet-printed with different resolutions, pre-dried at different temperatures, and subsequently annealed at 160 °C for 5 min*